\documentclass[epj]{svjour}

\usepackage{graphicx}
\usepackage{atlasphysics} 

\usepackage{cite}
\usepackage{./mcite}
\usepackage{url}
\usepackage{hyperref}

\newcommand{\ptl}{p_{\mathrm{T}}^\mathrm{lead}}
\def\TeV{\ifmmode {\mathrm{\ Te\kern -0.1em V}}\else
                   \textrm{Te\kern -0.1em V\,}\fi}%
\def\GeV{\ifmmode {\mathrm{\ Ge\kern -0.1em V\,}}\else
                   \textrm{Ge\kern -0.1em V\,}\fi}%
\def\MeV{\ifmmode {\mathrm{\ Me\kern -0.1em V}}\else
                   \textrm{Me\kern -0.1em V\,}\fi}%

\begin{document}




\title{
Measurements of underlying-event properties
using neutral and charged particles 
in $pp$ collisions
at $\sqrt{s}=900$~\GeV and $\sqrt{s}=7$~\TeV with the ATLAS detector at the LHC}

\author{The ATLAS Collaboration
}

\institute{ }

\abstract{
We present first measurements of charged and neutral particle-flow
correlations in $pp$
collisions using the ATLAS calorimeters.
Data were collected in 2009 and 2010
at centre-of-mass energies of 900~\GeV and
7~\TeV. Events were selected using a
minimum-bias trigger which required a charged particle in scintillation
counters on either side of the interaction point.
Particle flows, sensitive to the underlying event, 
are measured using clusters of energy in the ATLAS calorimeters, taking advantage
of their fine granularity. 
No Monte Carlo generator used in this analysis can accurately describe the measurements.
The results are independent of those based on charged particles 
measured by the ATLAS tracking systems and can be used 
to constrain the parameters of Monte Carlo
generators.
}


\maketitle
\section{Introduction}
All hard parton-parton interactions in $pp$ collisions 
are accompanied by additional processes which collectively 
contribute additional particles to those from the hard scatter
and which are termed the underlying event (UE). 
It is impossible to uniquely separate the UE from the hard scattering process 
on an event-by-event basis. However, observables can  be measured which are sensitive to 
properties of the UE. In order to make high-precision measurements, the UE must be modelled using 
phenomenological models in Monte Carlo generators \cite{r.field}.
Such models must be tuned to experimental data.
In the past, such studies have only been performed using 
tracks~\cite{CDF:mbpaper,dd:2010rm, CMS:UE,ATLAS:UE}.

Many physics processes to be studied with the  ATLAS detector~\cite{ATLAS:detector} require precision measurements
of jets and missing transverse energy obtained principally from the calorimeter system.
Therefore, it is important that the UE measurements are performed
using the same instrumental environment and reconstructed 
objects as those for the calorimeter-based measurements.
The fine granularity of the ATLAS calorimeter allows the definition of 
three-dimensional clusters of energy 
which  are closely associated with individual 
particles \cite{Lampl:1099735,Barillari:1112035}.  

\begin{figure}[htbp]
  \begin{center}
   \includegraphics[width=3.6cm]{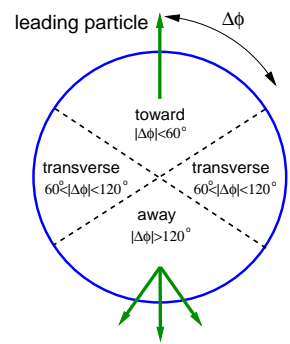}
   \caption{A schematic representation of 
    regions in the azimuthal angle $\phi$ with respect to the leading particle (shown with the arrow).
    In this analysis, the leading particle corresponds 
    to the cluster with the largest transverse momentum.}
    \label{fig:uediagram}
  \end{center}
\end{figure}

A study of the UE using charged-track densities was  recently performed by 
ATLAS \cite{ATLAS:UE}. The present paper extends this 
measurement by reconstructing particle densities using calorimeter clusters in 
the region which is most sensitive to the soft QCD processes 
responsible for the UE; the ``transverse'' region as 
shown in Fig.~\ref{fig:uediagram}. 
The azimuthal angular distance between a leading particle 
in transverse momentum ($\pT$)
and other particles is given by 
$\Delta\phi=\phi-\phi_{\,\mathrm{lead}}$, 
where $\phi$ is the azimuthal angle of a particle and $\phi_{\,\mathrm{lead}}$ is the 
azimuthal angle of the leading particle.  
The transverse region, defined
as $60^\circ<|\Delta\phi|<120^\circ$, 
is  most sensitive to the UE since it is  perpendicular to the axis
of hardest scattering,  approximated by the direction of the leading particle.
As is the 
case for charged particles, the number density of the clusters 
and their transverse energy density in this region are 
sensitive, discriminating observables for UE studies. 
These distributions are corrected for detector effects to give a measure of the 
particle activity in the UE and 
to provide new characteristics which can be used to tune models included in Monte Carlo generators.

The analysis using calorimeter clusters has several important features. 
Firstly, its results are sensitive to the entire hadronic final 
state, including neutral particles, which constitute about 
$40\%$ of all produced particles. Secondly, the analysis based  
only on calorimeter clusters 
has completely independent experimental uncertainties compared to 
the corresponding analysis \cite{ATLAS:UE} using charged particles.
Finally, as discussed earlier, 
since jet reconstruction is based almost 
entirely on energy deposition in the calorimeter, the results of this  UE analysis 
can be used directly to estimate the effect of 
the underlying event on any jet-based measurement.

\section{The ATLAS Detector}
The ATLAS detector~\cite{ATLAS:detector} at the Large Hadron Collider was 
designed to study a wide range of physics. It covers
almost the entire solid angle around the collision point with layers of tracking
detectors, calorimeters and muon chambers.

Charged tracks and vertices are reconstructed with the  inner detector which
consists of a silicon pixel detector, a silicon strip detector and a
transition radiation tracker, all  immersed in a 2 tesla magnetic field provided
by a superconducting solenoid.
For the measurements
presented in this paper, the high-granularity calorimeter systems are
of particular importance. The ATLAS calorimeter system provides fine-grained  
measurements of shower
energy depositions over a large range in
pseudorapidity\footnote{
The ATLAS reference system is a Cartesian right-handed co-ordinate system,
with the nominal collision point at the origin.
The anti-clockwise beam direction defines the positive $z$-axis,
while the positive $x$-axis is defined as pointing from the collision
point to the centre of the LHC ring and the positive $y$-axis points upwards.
The azimuthal angle $\phi$ is measured around the beam
axis, and the polar angle $\theta$ is the angle measured with respect to the $z$-axis. 
The pseudorapidity
is given by $\eta=-\ln\tan(\theta/2)$. Transverse momentum is defined relative to the beam axis.}. 
Electromagnetic calorimetry in the range $|\eta|<3.2$ is
provided by liquid argon (LAr) sampling calorimeters. This calorimeter system
provides measurements of the shower energy in up to four depth segments and  with transverse 
granularity that ranges from $0.003\times0.10$ to 
$0.05\times0.025$ in $\delta\eta \times \delta\phi$, depending on depth segment and rapidity. 
The hadronic calorimetry
in the range $|\eta|<$ 1.7 is provided by a steel/scintillator-tile 
sampling calorimeter.
This system provides measurements of the shower energy deposition
in three depth segments at a transverse granularity of typically $0.1\times0.1$.
In the end-caps ($|\eta|>1.5$), LAr technology is used for the
hadronic calorimeters that  match the outer $\eta$ limits of the end-cap
electromagnetic calorimeters. This system provides four measurements in depth
of the shower energy deposition at a transverse granularity of either
$0.1\times0.1$ ($1.5<|\eta|<2.5$) or $0.2\times0.2$ ($2.5<|\eta|<3.2$). The LAr forward
calorimeters provide both electromagnetic and hadronic energy measurements and 
extend the calorimeter coverage from $|\eta|=3.2$ to $|\eta|=4.9$.
To measure the energy of photons and
electrons, all calorimeter cells are initially calibrated to the
electromagnetic energy scale using 
test-beam data~\cite{CALOR:H1calib,Abat:2009zz,Adragna:2009zz,Aharrouche:2010zz,2009zp,2010dd}.

This analysis is based on the properties of
topological clusters in the calorimeter, which  represent an attempt to reconstruct
three-dimensional energy depositions
associated with individual particles~\cite{Lampl:1099735,Barillari:1112035}. 
The topological-cluster algorithm proceeds through the following steps.    
Nearest neighbours are collected around
seed cells, which are cells with an absolute signal greater than $4\sigma$
above the noise level \cite{CALOR:H1calib,Abat:2009zz,Adragna:2009zz,Aharrouche:2010zz,2009zp,2010dd}.
Then, neighbouring cells are collected into the cluster if the
absolute value of their signal significance is above a secondary seed
threshold of 2$\sigma$.
All surrounding cells are iteratively added to the cluster until no  further
secondary seeds are among the direct neighbours. A final analysis of
the resulting cluster looks for multiple local signal maxima; in the case 
of more than one maximum in a given cluster, it is split into
smaller clusters along the signal valleys between the maxima. 

The analysis
presented in this paper uses calibrated topological clusters~\cite{Barillari:1112035}.
The clusters are classified as related
to electromagnetic or hadronic energy deposits, using detailed information on the cluster
topology. 
Then, the reconstructed cluster energy is corrected for the non-compensating nature
of the ATLAS calorimeter and for inactive material.

\section{Data selection}
\label{data}

The data taken at $\sqrt{s}=900$~\GeV were collected during 6--15 December 2009. 
During this running period, there
were approximately 3\% non-functional channels in the tile hadronic
calorimeter and approximately 1\% non-functional channels
in the LAr calorimeters \cite{2009zp,2010dd}.
For an integrated luminosity of 7~$\mu \mathrm{b}^{-1}$,  
a total of $455$ thousand  events were collected from colliding proton bunches in which
a minimum-bias trigger recorded one or 
more hits in the scintillators
on either side of the detector. 

The events to be analysed were selected using 
a procedure identical to that described in Refs.~\cite{ATLAS:UE,Aad:2010rd}.
Events were required to have  
a primary vertex which is  
reconstructed using at least two tracks with transverse
momenta  $\pT>100$~\MeV and  
a transverse distance of closest approach with respect to the 
beam-spot position \cite{ATLAS-CONF-2010-069} of less than $4$~mm.

This analysis uses topological clusters with 
$\pT>0.5$ \GeV and $|\eta|<2.5$  in order to have the same kinematic 
range\footnote{
The topological clusters are treated as massless particles,
and we choose to refer to both clusters and stable particles
in terms of $\pT$.
The same symbol $\pT$ is
also used to represent the track transverse momentum.}
as for the previous UE study based on tracks~\cite{ATLAS:UE}.
Additional selection criteria were applied to select good-quality clusters:
1) to reject the cosmic and
noise background, the leading cell energy of the cluster is required to be 
less than 90\% of the cluster
energy;
2) the energy sampling maximum 
must be in a calorimeter region with good calibration; 
3) the fraction of energy in the cluster associated with problematic 
cells (or dead cells where the energy contribution is obtained by
interpolation from adjacent cells) should be less than 50\%.

Data at $\sqrt{s}=7$~\TeV  were collected between 30 March and 27 April 2010. 
Only a fraction of the 7~\TeV data, corresponding to an integrated 
luminosity of about 230~$\mu \mathrm{b}^{-1}$, was used.
In total, about 7.7 million events were analysed. 
Event selection was similar to that for the 900 GeV data, but included the additional
requirement of a single primary vertex \cite{ATLAS:UE,Aad:2010rd} to 
remove events containing more than one $pp$ interaction. 

\section{Monte Carlo simulation}

The QCD predictions for the hadronic final state
in inelastic $pp$ collisions are based on 
several Monte Carlo generators. 
The PYTHIA 6.4 Monte Carlo generator~\cite{Sjostrand:2006za} is
used as the primary
generator for comparisons with the data.
The MC09 tune \cite{ATL-PHYS-PUB-2010-002} 
of this model was performed by ATLAS. It uses
the $\pT$-ordered parton shower
with the MRST LO$^{*}$  parton-density function~\cite{Sherstnev:2007nd},
followed by fragmentation 
into final-state particles using the Lund string model \cite{Andersson:1983ia}. 
The parameters of this generator were adjusted to describe charged-particle multiplicity 
distributions in
minimum-bias events measured at $\sqrt{s}=630$~\GeV 
and $\sqrt{s}=1.8$~\TeV in $p\bar{p}$ 
collisions  \cite{Abe:1988yu}. 
Diffractive processes are not included in the simulation for the main
samples, but were used for systematic checks (Sect.~\ref{sys}).
In addition to the MC09 tune, 
the following two PYTHIA parameter sets are also used: (1) the Perugia0 
set \cite{Skands:2009zm} in which
the soft-QCD part is tuned using only minimum-bias data from the Tevatron 
and CERN $p\bar{p}$  colliders;
(2) the DW\cite{DW}  PYTHIA tune, which uses virtuality-ordered 
showers and was derived to describe the
CDF Run II underlying event and Drell-Yan data.

The data are also compared to the PHOJET Monte Carlo 
generator~\cite{Engel:1999xb}, which includes a simulation of the diffractive
component. This generator is based on the two-component Dual Parton Model
which includes soft hadronic processes described by Pomeron exchange 
and semi-hard processes
described by perturbative parton scattering. 
The description of the fragmentation is the same as in the PYTHIA generator. 

In addition, the HERWIG 
Monte Carlo generator~\cite{Corcella:2000bw,Corcella:2002jc} was used for 
comparisons with the data. 
This generator has similar matrix-element calculations as PYTHIA, 
but uses the cluster fragmentation model to hadronise partons into hadrons.
HERWIG is interfaced with the JIMMY model \cite{Butterworth:1996zw} 
in order to describe multiple parton interactions.

Monte Carlo events were processed through the ATLAS detector simulation 
program \cite{Marshall:1136205},
which is based on {\sc Geant4} \cite{Agostinelli2003250}. They were 
reconstructed using the same trigger and event selection as for the data.
The size and position of the collision beam-spot and 
the detailed description of detector conditions during the data-taking runs were included in the simulation.

Monte Carlo events after the detector simulation program were used for 
correcting the data to the stable-particle level defined as follows.  
The PYTHIA MC09 is used to generate
the primary
samples for unfolding the effects of the detector.
Monte Carlo stable particles are selected
if their mean lifetimes are larger than $3\cdot 10^{-11}$ seconds.  
Neutrinos are excluded from consideration.
According to this definition, $K^{0}_{S}$,  $\Lambda$ and 
$\Sigma^{\pm}$ are among those treated as stable particles.
This definition allows a direct comparison between the results of previous 
track-based studies~\cite{ATLAS:UE} and the present measurement.

\section{Properties of calorimeter clusters}
\label{sec:prop}

\begin{figure*}
\begin{minipage}[b]{0.48\linewidth}
\centering
\includegraphics[scale=0.3]{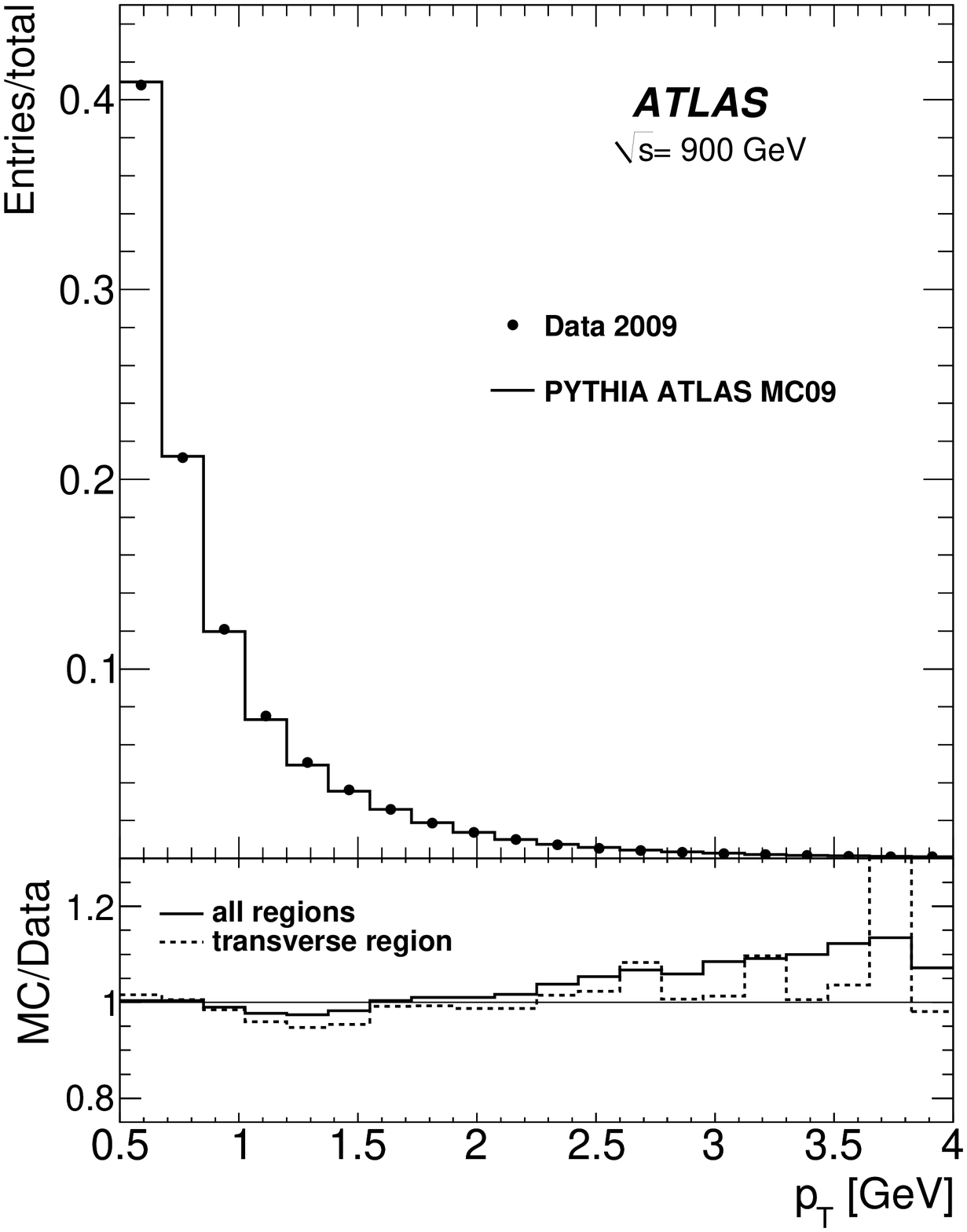}
\end{minipage}
\hspace{0.3cm}
\begin{minipage}[b]{0.48\linewidth}
\centering
\includegraphics[scale=0.3]{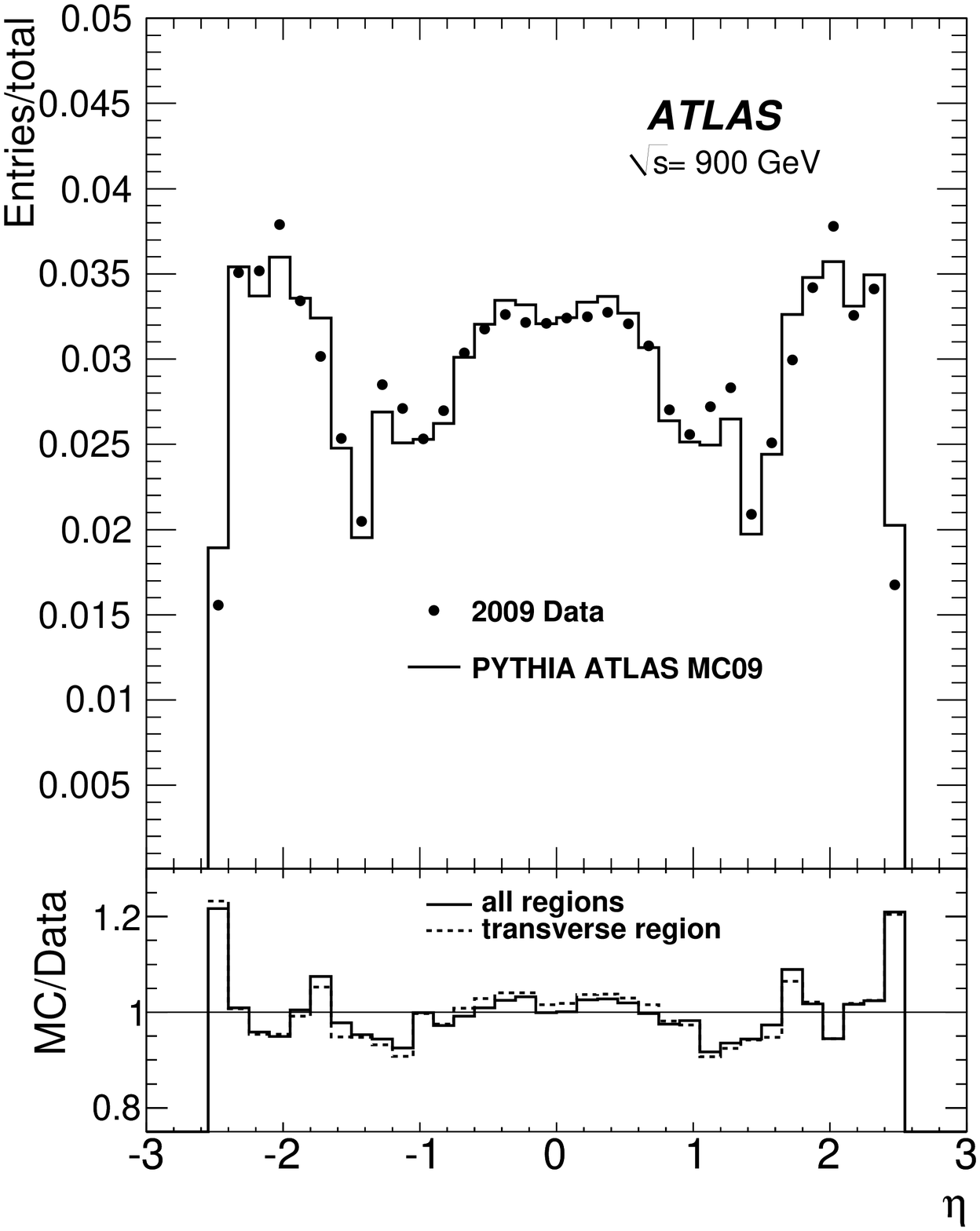}
\end{minipage}
\begin{picture} (0.,0.)
   \setlength{\unitlength}{1.0cm}
    \put (-13.4, 0.){(a)}
    \put (-4.3,  0.){(b)}
 \end{picture}
\vspace*{0.5cm}
\caption{
A comparison between uncorrected data and the Monte Carlo simulation for topological cluster
$\pT$ (a) and $\eta$ (b) for $pp$ collisions at $\sqrt{s}=900$~\GeV.
The ratio plots show
the inclusive sample (solid lines) and the transverse region (dashed lines).
}
\label{fig:900gev}
\end{figure*}

\begin{figure*}
\begin{minipage}[b]{0.48\linewidth}
\begin{center}
\includegraphics[scale=0.3]{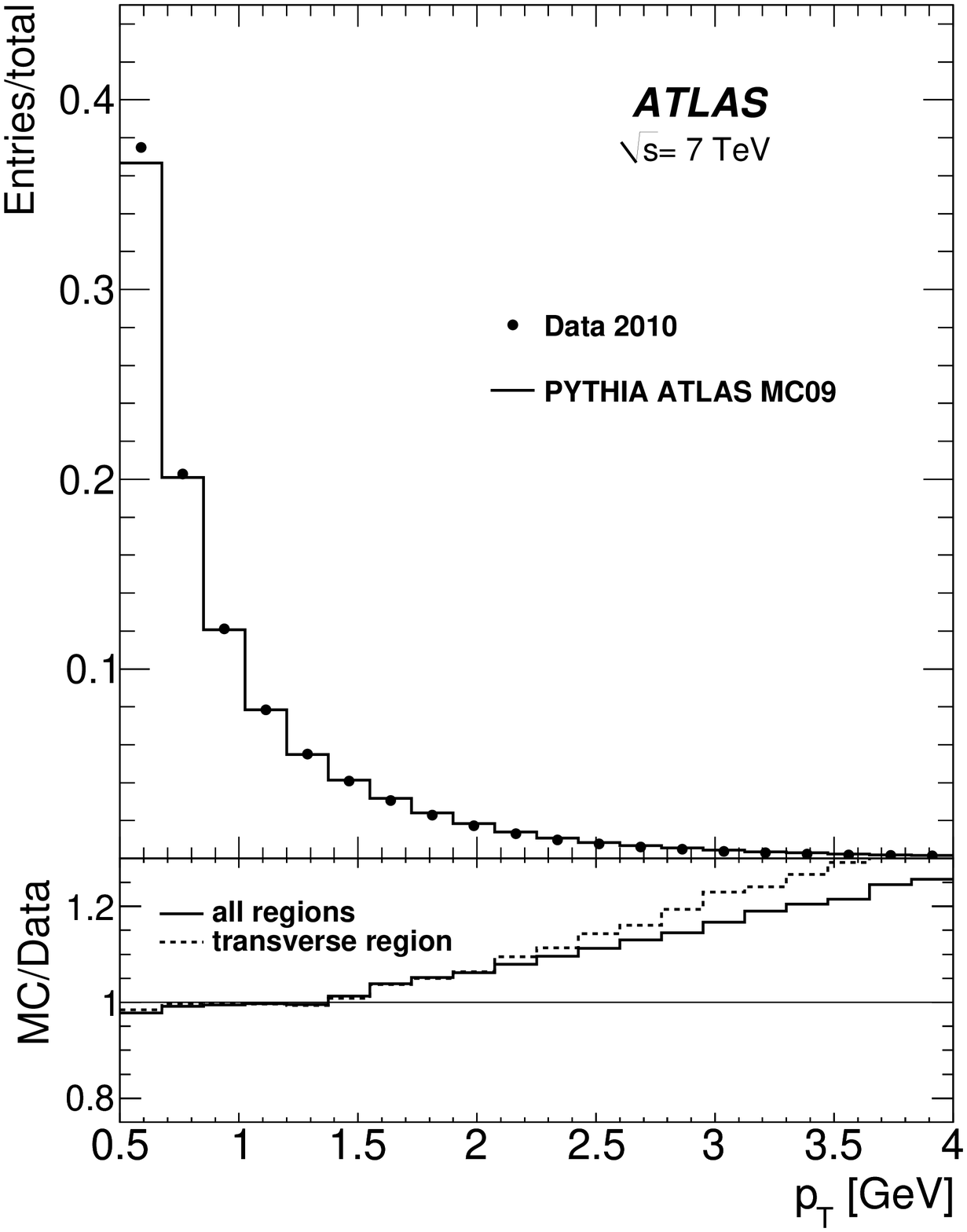}
\end{center}
\end{minipage}
\hspace{0.3cm}
\begin{minipage}[b]{0.48\linewidth}
\centering
\includegraphics[scale=0.3]{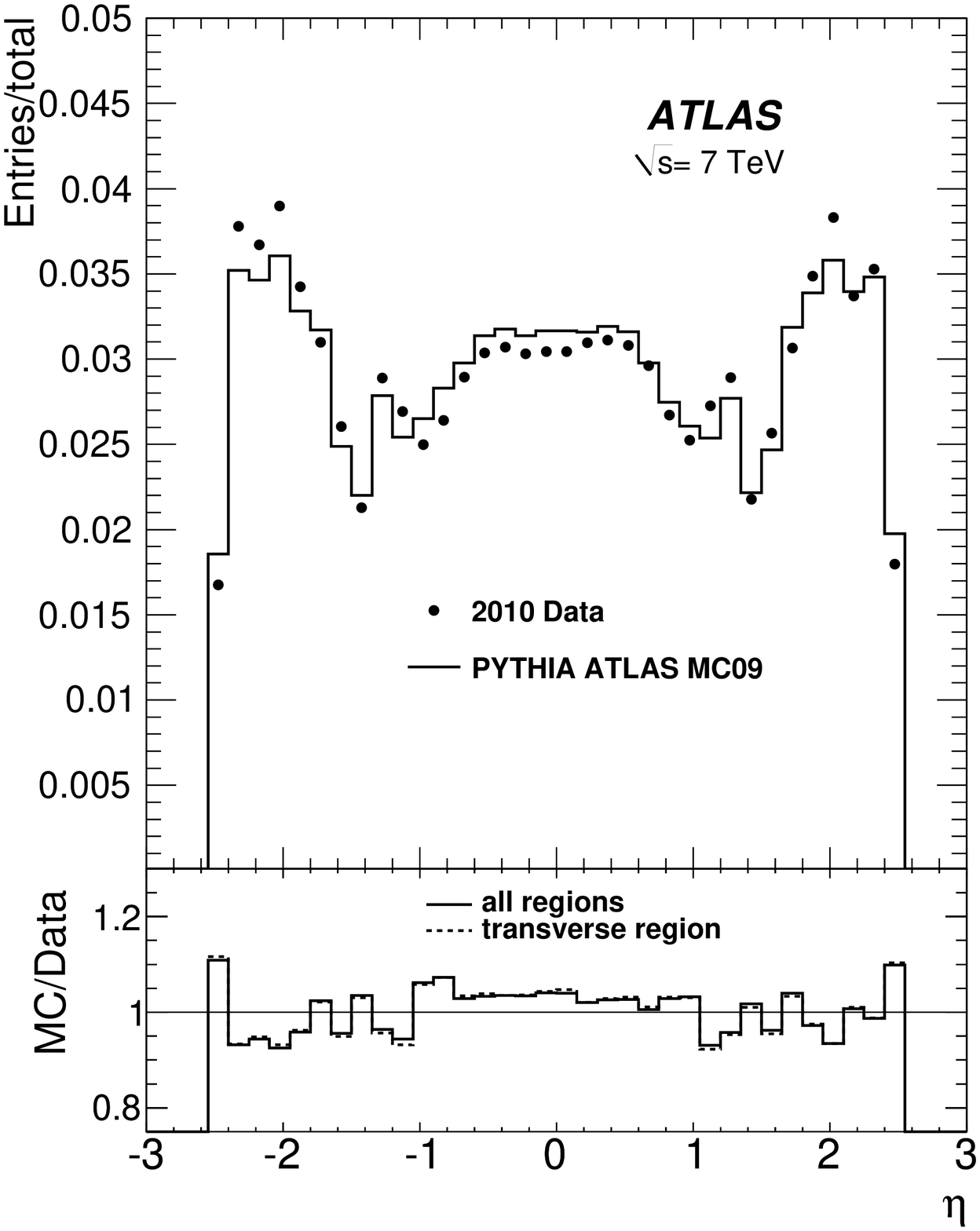}
\end{minipage}
\begin{picture} (0.,0.)
   \setlength{\unitlength}{1.0cm}
 \put (-13.4, 0.){(a)}
 \put (-4.3,  0.){(b)}
 \end{picture}
\vspace*{0.5cm}
\caption{
A comparison between uncorrected data and the Monte Carlo simulation for topological cluster
$\pT$ (a) and $\eta$ (b) for $pp$ collisions at $\sqrt{s}=7$~\TeV.
The ratio plots show
the inclusive sample (solid lines) and the transverse region (dashed lines).
}
\label{fig:7tev}
\end{figure*}

Figures~\ref{fig:900gev} and \ref{fig:7tev} show the distributions of $\pT$ and $\eta$ for topological clusters
in data and simulated PYTHIA MC09 events at $\sqrt{s}=900$~\GeV and  $\sqrt{s}=7$~\TeV, 
respectively. The distributions in each case are normalised to the number of entries. In addition, 
the ratio plots show
the ratio of simulation to data in the transverse region alone.  
The figures show overall good agreement between the data and the  PYTHIA MC09 tune, with
$20\%$ discrepancies in some phase-space regions. 
While not shown in these figures, the Perugia0 tune
agrees with the data to a similar extent.
The contribution of the discrepancy in the high-$\pT$ tail  
is expected to be small on particle densities measured at $\pT>0.5$ \GeV, 
and it was taken into account
using re-weighting as described below.
The observed differences between the data and the PYTHIA MC09 event sample for 
the $\eta$ distributions are addressed in the studies of
systematic uncertainties.

\begin{figure*}
   \begin{center}
   \includegraphics[width=14cm]{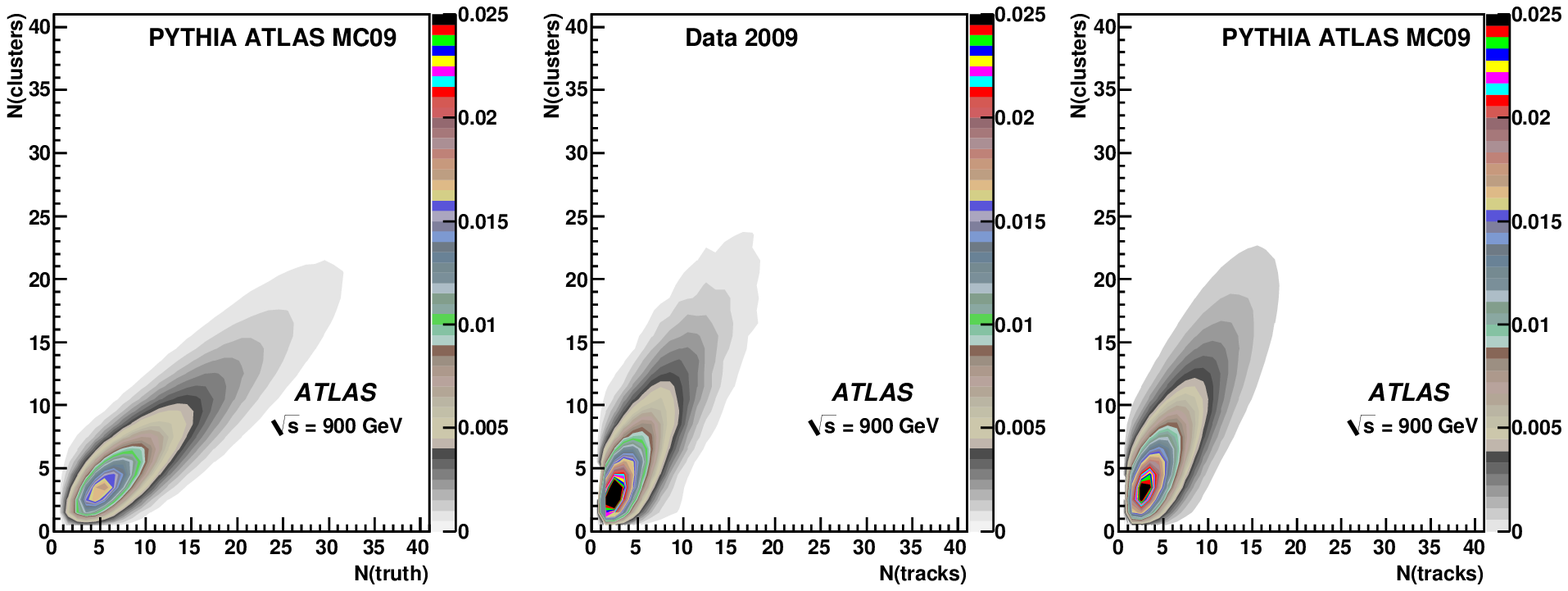}
   \begin{picture} (0.,0.)
   \setlength{\unitlength}{1.0cm}
   \put (-12.2,-0.05){(a)}
   \put (-7.25,-0.05){(b)}
   \put (-2.8,-0.05){(c)}
 \end{picture}
\vspace*{0.5cm}
  \caption{
     ATLAS data at $\sqrt{s}=900$~\GeV:
     The correlations between the multiplicities of, (a)  topological clusters ($N(\mathrm{clusters})$) and stable
     particles ($N(\mathrm{truth})$) from simulated $pp$ interactions, (b) topological clusters and primary
     reconstructed tracks ($N(\mathrm{tracks})$) from $pp$
     interactions, and (c) topological clusters and primary reconstructed tracks from simulated $pp$ interactions.
     Inelastic events generated by PYTHIA MC09 (without diffraction) passed through the selection were used to produce the plots (a) and (c). }
    \label{fig:correlations900}
  \end{center}
\end{figure*}

\begin{figure*}
   \begin{center}
   \includegraphics[width=14cm]{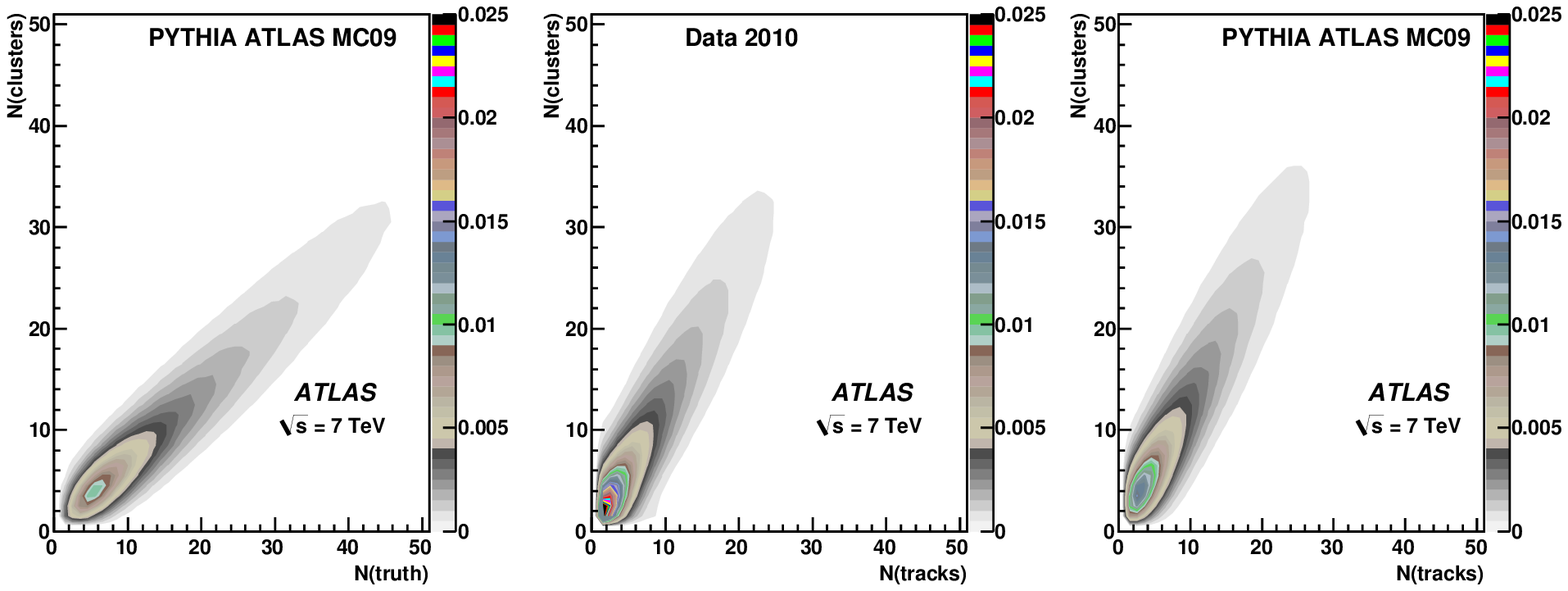}
   \begin{picture} (0.,0.)
   \setlength{\unitlength}{1.0cm}
   \put (-12.2,-0.05){(a)}
   \put (-7.25,-0.05){(b)}
   \put (-2.8,-0.05){(c)}
 \end{picture}
\vspace*{0.5cm}
  \caption{
     ATLAS data at $\sqrt{s}=7$~\TeV: The correlations between the multiplicities of, (a)  topological clusters  ($N(\mathrm{clusters})$) and stable
     particles ($N(\mathrm{truth})$) from simulated $pp$ interactions, (b) topological clusters and primary reconstructed tracks ($N(\mathrm{tracks})$) from $pp$
     interactions, and (c) topological clusters and primary reconstructed tracks from simulated $pp$ interactions.
     Inelastic events generated by PYTHIA MC09 (without diffraction) passed through the selection were used to produce the plots (a) and (c).
     }
    \label{fig:correlations7}
  \end{center}
\end{figure*}

Figures~\ref{fig:correlations900}(a) and  \ref{fig:correlations7}(a) show the multiplicity of 
topological clusters, with $\pT>0.5$~\GeV and $|\eta| <2.5$,
versus the number of stable particles (charged and neutral) in simulated 
events for $\sqrt{s}=900$~\GeV and  $\sqrt{s}=7$~\TeV. A strong correlation is observed between the
number of topological clusters and the number of stable particles, indicating 
that  clusters are a good representation of the particle activity in inelastic $pp$ events.

Figures~\ref{fig:correlations900}(b) and  \ref{fig:correlations7}(b)  show the correlation
between the number of topological clusters
and the number of primary tracks
selected in the same way as in the track-based studies~\cite{ATLAS:UE,Aad:2010rd}. 
These figures also show a strong correlation. 
The Monte Carlo simulation shown in Fig.~\ref{fig:correlations900}(c) and ~\ref{fig:correlations7}(c) 
reproduces these distributions 
well: the means and the root-mean-square deviations of one-dimensional projections of these distributions 
agree with the Monte Carlo simulation within less than one percent for $N(\mathrm{tracks})>4$. 
For events with a lower number of tracks, the data show a smaller mean value of 
the projection onto the $x$-axis
than seen in the PYTHIA MC09 simulation.  
This is attributed to the absence of 
diffraction in the generated samples.

A Monte Carlo simulation study based on PYTHIA MC09 indicates that
the probability that a second particle lies within 
$\sqrt{\delta \eta^2+\delta \phi^2}<0.2$ of a first in 
the selected inelastic $pp$ events is below $1\%$.
This simplifies the present analysis since there is negligible potential bias due 
to cluster overlap.

For the UE studies based on topological clusters, a good position measurement is required.
The quality of the
position reconstruction of the clusters was studied
by comparing the impact point of charged particles with the associated
cluster position in the calorimeter.
Charged particles are deflected in the magnetic field of the solenoid. Their trajectories
are extrapolated to the calorimeter using a Monte Carlo simulation which includes a detailed field map as well
as the effect of the material in front of the calorimeter.
The Monte Carlo simulation describes
the topological-cluster positions relative to the positions of the extrapolated tracks on the 
surface of the LAr calorimeter within the granularity of its second 
layer ($0.025\times0.025$ in $\delta\eta \times \delta\phi$).

\begin{figure}[ht]
\centering
\includegraphics[scale=0.4]{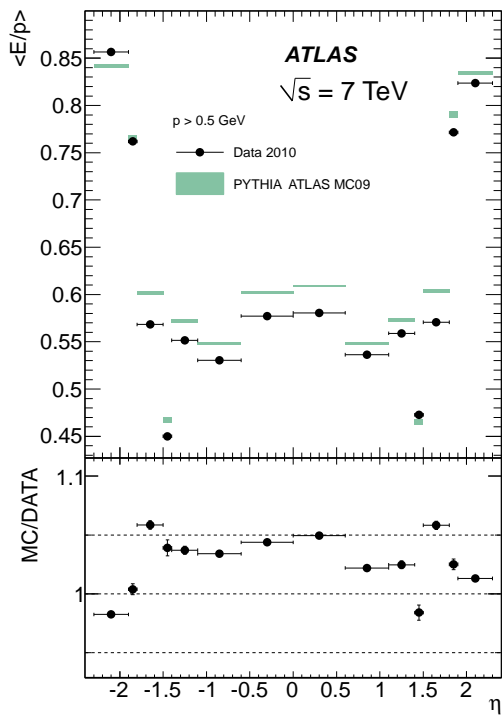}
\caption{The average $E/p$ in different $\eta$ bins for isolated
topological clusters matched to charged tracks
in inelastic $pp$ events at $\sqrt{s}=7$~\TeV  for
track momentum $p$ larger than 0.5~\GeV.
A similar level of agreement between data and the Monte Carlo simulation was obtained for the $\sqrt{s}=900$~\GeV data (not shown).
}
\label{fig:lcw}
\end{figure}

As the correction for detector effects is based on the Monte Carlo simulation, an
essential issue is the accuracy with which the simulation reproduces the
energy reconstruction in the calorimeter. 
For charged particles, 
the energy scale was studied \cite{Barberio:1271712,ATLAS-CONF-2010-017} using 
isolated tracks by 
extrapolating tracks to the calorimeter surface and matching them to 
topological clusters.
The average value of the ratio $E/p$ was reconstructed,  where
$E$ is the cluster energy in the calorimeter and $p$ is the track momentum.
Figure~\ref{fig:lcw} shows the average
response $\langle E/p\rangle$ for  calibrated topological clusters 
as a function of $\eta$ for tracks with $p>0.5$~\GeV.
The data and PYTHIA MC09 agree within $5\%$ in most $\eta$ regions, while discrepancies
increase in the transition region ($1.5<|\eta|<1.8$)
between barrel and end-cap. 

To estimate the relative energy-scale
uncertainty, the double ratio $\langle E/p\rangle_{MC}/\langle E/p\rangle$ was calculated, 
where the ratio $\langle E/p\rangle_{MC}$ was determined from the Monte Carlo simulation. 
The double ratio as a function of $\eta$ is shown in Fig.~\ref{fig:lcw}~(bottom).
The double-ratio distributions were measured  
for a wide range of track momenta and $\eta$ as
described in Refs.~\cite{Barberio:1271712,ATLAS-CONF-2010-017}.  

\begin{figure*}[htbp]
  \begin{center}
   \includegraphics[width=12cm]{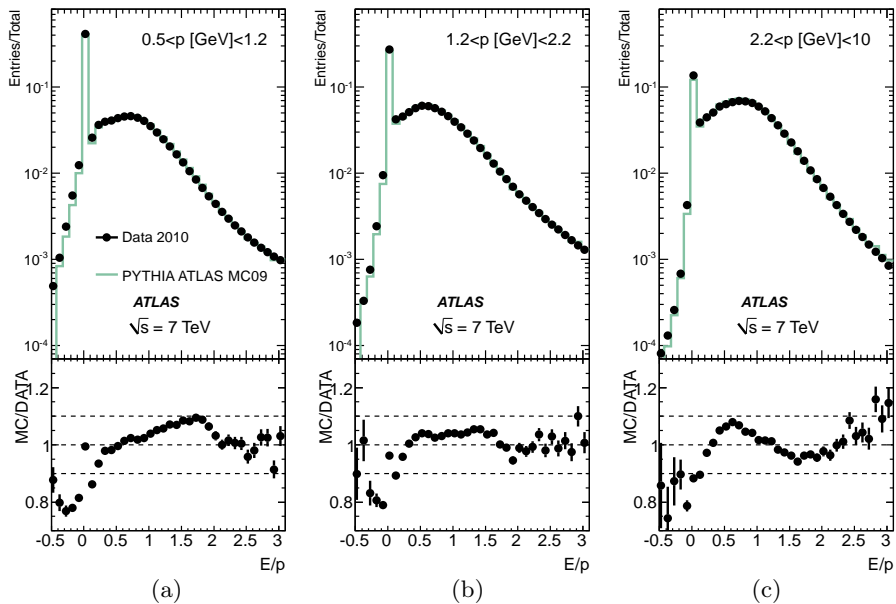}
     \begin{picture} (0.,0.)
   \setlength{\unitlength}{1.0cm}
   \put (-10.2,0.1){(a)}
   \put (-6.25,0.1){(b)}
   \put (-2.3,0.1){(c)}
 \end{picture}
   \caption{$E/p$ distributions at $\sqrt{s}=7$~\TeV for topological clusters
    matched to tracks in several bins of track momentum: (a) $0.5<p<1.2$ \GeV, (b) $1.2<p<2.2$ \GeV and (c) $2.2<p<10$ \GeV.
     The peak at zero corresponds to the 
events without a good match between a topological cluster and a track.
    A similar level of agreement between data and Monte Carlo was obtained for the $\sqrt{s}=900$~\GeV data (not shown).}
    \label{fig:clusters_reso_EoverP}
  \end{center}
\end{figure*}

The comparison  between data and Monte Carlo predictions for the shapes of 
the $E/p$ distribution is shown 
in Fig.~\ref{fig:clusters_reso_EoverP}. The peak at zero 
corresponds to isolated tracks that have no associated cluster  
in the calorimeter. These are predominantly due to hadronic interactions
in the material in front of the calorimeter~\cite{ATLAS-CONF-2010-017}. 
The contribution of the discrepancies observed for $E/p=0$ between the data and PYTHIA MC09
to uncertainties on the reconstruction efficiencies of topological clusters is below $1\%$. This effect 
was taken into account as described in Sect.~\ref{sys}.  
More details 
on the energy scale of topological clusters can be found 
in Refs.~\cite{Barberio:1271712,ATLAS-CONF-2010-053}.

The energy scale for electromagnetic clusters was estimated
using the $\pi^0$ peak reconstructed  in inelastic $pp$ events.
The selection criteria for calibrated 
topological clusters were the same as for the present analysis.
The $\pi^0$ peak positions for data and PYTHIA MC09 agree 
within $3\%$ for all $\eta$ regions.

Correction of the observed distributions
to the particle level requires a reliable description of the cluster
multiplicity distribution by the simulated event sample. 
This was studied
by examining cluster multiplicities in bins of track multiplicity using 
projections of the two-dimensional distributions
shown in Figs.~\ref{fig:correlations900} and
\ref{fig:correlations7}.
The observed differences are propagated into the systematic uncertainties
as discussed in  Sect.~\ref{sys}.

For the UE studies, the so-called ``leading''
clusters, i.e.
clusters with the largest transverse momenta, $\ptl$, are used
to define an event orientation. Such clusters are typically
inside the most energetic jets. To verify this, 
jets were reconstructed with the anti-$k_t$ algorithm \cite{Cacciari:2008gp}
with  a distance parameter of 0.4, a minimum $\pT$ requirement of $5$~\GeV and $|\eta|<2.5$. Then,  
the distance in $\eta$--$\phi$  
between the leading topological cluster  and the centre of the leading jet was calculated.
It was shown that, in the vast majority of cases, the leading cluster
is inside a leading jet with only a small fraction $(\simeq 10\%)$ of clusters 
opposite the leading jet in $\phi$.
This feature is well reproduced by the  Monte Carlo simulation.

To verify the Monte Carlo performance for $\ptl$, the ratio of $\ptl$ of topological clusters 
to $\ptl$ of primary tracks
was reconstructed. The agreement between the data and the PYTHIA MC09 tune for such distributions 
was found to be within $\pm 5\%$ in most regions,
while discrepancies at the level of $20\%$ were found  for the tails of the ratio distributions.
The impact of such discrepancies in the simulation of the $\ptl$ resolution 
on the final measurement has been estimated as discussed in Sect.~\ref{sys}.

Monte Carlo studies show that 
the rate of events in which a low-$\pT$ 
particle  is reconstructed as a high-$\pT$ cluster is not negligible.
This results in a low purity for topological clusters at high $\ptl$.
Therefore,  the analysis was performed for  leading topological clusters with
transverse momenta less than $8$~\GeV ($14$~\GeV) for the $\sqrt{s}=900$~\GeV \\ (7~\TeV) data in order
to limit this effect and 
to ensure that the reconstruction purity even
at the highest transverse momenta considered is larger
than $50\%$.

\section{Measured observables and correction procedure}
\label{corr}

Following earlier track-based analyses~\cite{ATLAS:UE},
particle densities are studied 
as a function of the distance $\Delta\phi$ in the 
azimuthal angle between the leading cluster and all other clusters in an 
event, and as a function of $\pT$ of the leading cluster in the event. 
The scalar $\pT$ sum for stable particles per unit area in $\eta$--$\phi$ in 
the transverse region is also presented. This provides complementary information to 
that which can be obtained from the particle densities.

The particles and clusters are required to have $\pT>0.5$~\GeV and $|\eta| < 2.5$.
Clusters are selected if they pass the criteria described in Sect.~\ref{data}.
The measured observables at the particle and detector levels are:

\begin{itemize}
\item $\ptl$
- Transverse momentum of the stable  particle with maximum $\pT$ in the event. At the detector level,
this corresponds to the transverse momentum of the selected topological 
cluster with maximum $\pT$  in the event. 

\item $d\langle N \rangle/d \Delta \phi$
- The average number of particles as a function of the azimuthal-angle difference between
the leading particle and other particles in an event.
The leading particle at
$\Delta \phi =0$  is excluded from this distribution.   
At the detector level, it corresponds to the
mean number of selected topological clusters as a function of the azimuthal-angle 
distance between
the leading topological cluster and other clusters in an event. 
This density \cite{ATLAS:UE} is defined per unit of pseudorapidity 
as $N / (N_{ev}\cdot (\eta_{\mathrm{max}}-\eta_{\mathrm{min}}) )$,
where $N$ is the number of entries in $\Delta \phi$ bins, 
$\eta_{\mathrm{max}}-\eta_{\mathrm{min}}=5$ represents the full pseudorapidity range, and
$N_{\mathrm{ev}}$ is the number of events selected by requiring a particle with
$\ptl$ above the specified value.

\item $\langle d^{2}N/d\eta d\phi\rangle$
- Mean number of stable particles per unit area in $\eta$--$\phi$. At the detector level, this corresponds to the
mean number of selected topological clusters  per unit area in $\eta$--$\phi$. This density is
measured as a function of $\ptl$ \cite{ATLAS:UE}.   

\item $\langle d^{2}\sum \pT/d \eta d\phi\rangle$
- Mean scalar $\pT$ sum for stable particles per unit area in  $\eta$--$\phi$. At the detector level, this corresponds to
the mean scalar $\pT$ sum for selected topological clusters  per unit area in $\eta$--$\phi$. 
This quantity is defined following the convention used in the previous ATLAS publication~\cite{ATLAS:UE}.
\end{itemize}

A bin-by-bin correction procedure is used to correct the observed distributions
to the stable-particle level. The correction factors 
$$
C =
\frac{\mathcal{A}^{\mathrm{gen}}}{\mathcal{A}^{\mathrm{det}}},
\label{cor1}
$$
are evaluated separately for each observable.
In the above expression, $\mathcal{A}^{\mathrm{gen}}$ is  calculated
at the stable-particle  level  of PYTHIA MC09 and  $\mathcal{A}^{\mathrm{det}}$ 
is calculated after full detector simulation and reconstruction.
The corrected value for an observable is  found by multiplying
its measured value by the relevant  correction factor $C$.
These factors correct the data to the stable-particle level and include the
effects of event selection,  
reconstruction efficiency, bin migrations   
and smearing, including the case when the leading particle
is mis-identified and 
a cluster corresponding to a sub-leading particle is used to define
the event orientation and $\ptl$.   

The bin-by-bin correction depends on the choice of the Monte Carlo event generator.
This affects the efficiency correction (mainly due to 
variations in particle types) and the purity (different 
stable-particle level distributions have different
fractions  of poorly reconstructed objects in each bin as well as different bin
migrations). To reduce the  model dependence of the correction procedure, 
bin-by-bin migrations were minimised by using bin sizes larger than the 
reconstruction resolutions for the distributions presented.
In  addition, the analysis was restricted to the $\ptl$ ranges where the 
purity of leading clusters is above $50\%$ (see Sect.~\ref{sec:prop}).

The bin-by-bin correction factors for the particle densities 
typically have values of around $1.3$ and do not exceed $1.4$.
The largest single contributor 
is the reconstruction inefficiency of topological clusters, which leads to 
a bin-by-bin correction factor of approximately $1.2$ on average and  
has a maximum value of $1.3$ at low $\pt$. The other 
significant contributor is 
the event reorientation which results from inefficiency of the reconstruction
of the leading topological cluster in an event. This causes bin migrations,
which were studied by replacing the leading cluster $\ptl$ by the leading track $\ptl(\mathrm{track})$,
for which the efficiency is known to be high \cite{ATLAS:UE}.   
The bin-by-bin corrections for the average scalar $\pT$ sum
have a maximum value of  $1.5$ for low $\ptl$ and decrease to $1.3$ for $\ptl>6$ \GeV.

To study the contribution from diffractive events,
the PYTHIA~\cite{Sjostrand:2006za} and PHOJET~\cite{Engel:1999xb} Monte Carlo generators were used. 
Non-diffractive inelastic $pp$ events were mixed
with single and double diffractive events in accordance with
the corresponding generator cross-sections for such processes.
The diffractive contribution was found to be below $1\%$  for the $d\langle N \rangle/d \Delta \phi$ densities at $\ptl>1$~\GeV
in PYTHIA, and almost entirely concentrated at low multiplicities (fewer than four topological clusters).
The contribution of diffractive events is larger for $\langle d^{2}N/d\eta d\phi\rangle$ and  
$\langle d^{2}\sum \pT/d \eta d\phi\rangle$  measured at $\ptl<3$~\GeV, but becomes negligible for $\ptl>3$~\GeV. 
Diffractive contributions are higher for PHOJET,  but their
contribution was found to be smaller than the systematic uncertainties on  
the final measurements. No attempt
to subtract diffractive events from the final measurements was made.

\section{Systematic uncertainties}
\label{sys}

The systematic uncertainties on the measured distributions 
were determined by changing the selection or the analysis procedure and
repeating the analysis. The largest uncertainties are 
described below:  

\vspace{0.2cm}

\begin{itemize}

\item[$\bullet$]

The following procedure was used to estimate the effect of the 
relative energy-scale uncertainty on the final measurements. 
The double ratio 
$\langle E/p\rangle_{MC} /\langle E/p\rangle$ was calculated for isolated single particles as described in Sect.~\ref{sec:prop}.
The effect of the energy-scale uncertainty on the measured densities was found by  
decreasing and increasing the $\pT$ of
topological clusters in the Monte Carlo simulation, keeping the same cluster $\pT$ in the data. 
The magnitude of the variation was set by the value of the double ratio 
calculated  in a grid defined in $\eta$ and $p$. 
To simplify the calculation of the systematic uncertainties, a common variation was used for all 
topological clusters independent of their origin (hadronic or electromagnetic).
The effect of the energy-scale uncertainty is
significantly larger than that due to the event selection (including
trigger) \cite{Aad:2010rd}.

\item[$\bullet$]
The dependence of the bin-by-bin corrections on the
detector material description was estimated by recalculating the corrections
using two further samples: one with an extra $10\%$ of material in the tracking
system, and the other with $\sim 15\%$  additional material in the region $|\eta|>2$.

\item[$\bullet$]
The physics-model dependence of the bin-by-bin corrections was estimated 
using the Perugia0 tune \cite{Skands:2009zm} instead of PYTHIA MC09. This uncertainty was symmetrised.

\item[$\bullet$]
A comparison of multiplicities of topological clusters in bins
of track multiplicities indicated some discrepancy between data and Monte Carlo for 
events with low track multiplicities (see Sect.~\ref{sec:prop}).  
To estimate a systematic uncertainty to account for this discrepancy, the bin-by-bin acceptance 
corrections were calculated after re-weighting the PYTHIA MC09 detector-level distributions.
For this, cluster multiplicity distributions were  measured in bins of  
track multiplicity and re-weighting factors
were calculated by taking the ratio of the above distribution in data and PYTHIA MC09.  
The re-weighting procedure also addresses the uncertainties on 
the noise description used in the Monte Carlo simulation and 
other effects related to the cluster-reconstruction efficiencies.   

\item[$\bullet$]
A systematic uncertainty was estimated to account for differences in the $\pT$ resolution 
of leading topological clusters in data compared to the Monte Carlo expectation.
Discrepancies in the tails of the distributions of $\ptl(\mathrm{clusters})/\ptl(\mathrm{tracks})$ 
were used to extract weighting factors, which were then used
to recalculate the acceptance corrections. 
\end{itemize}
\vspace{0.2cm}

Table~\ref{tab_system} shows 
the values of the systematic uncertainties discussed above as a percentage of the measured values.  
Only the largest values are shown for the bins with the most significant effect
from the selection variations or change in the experimental procedure.

In addition to these uncertainties, 
the following systematic variations were also included:   
1) in order to reduce the contribution from
diffractive events, the measurement was repeated  
after removing events with
fewer than four clusters; 
2) the positions of cluster centres 
in $\eta$ and $\phi$ were shifted by the size of one cell;
3) an alternative model (FTFP-Bertini) for the hadronic-shower simulation in 
{\sc Geant4} was used to extract the
correction factors; 4) the calorimeter transition region 
of $0.94<|\eta|<1.06$,  which is not well described by the Monte Carlo 
simulation was removed in the data and in the simulated  PYTHIA MC09 sample.
These variations each give systematic uncertainties below $2\%$,
with the exception of that for diffractive events which indicate a $7-10\%$ systematic 
uncertainty for the $\langle d^{2}N/d\eta d\phi\rangle$ and $\langle d^{2}\sum \pT/d \eta d\phi\rangle$ 
densities measured at $\ptl<3$~\GeV.
As an additional systematic check, the measurement was also repeated
using topological clusters at the electromagnetic energy scale and similar differences between data and Monte Carlo
simulations were observed.   

The overall systematic uncertainty was  
determined by adding the  above  uncertainties in quadrature.

\begin{table*}
\centering
   {\small
   \begin{tabular}{|c|lll|}
\hline 
Check  & $d \langle N\rangle  / d\Delta\phi$ &  $\langle d^{2}N/d\eta d\phi\rangle$  &  $\langle d^{2}\sum p_{T}/d \eta d\phi\rangle$  \\ \hline \hline 
Energy scale & $\pm4.3\%$            & $\pm4\%$    &   $\pm5.6\%$  \\ 
Additional material    & $+3.5\%$   &  $+3\%$   &  $+3.6\%$  \\ 
Model dependence   & $\pm3.5\%$       &  $\pm5\%$   &   $\pm4.5\%$   \\ 
Multiplicity reweighting  & $\pm4.5\%$ &  $\pm10\%$   &   $\pm11\%$   \\ 
Resolution   reweighting  & $\pm0.4\%$ &  $\pm6\%$   &   $\pm6\%$   \\ \hline 
\end{tabular} 

   }
\vspace*{0.5cm}
\caption{
A summary of the most important systematic uncertainties. 
The table lists the values of contributions from different groups 
of systematic checks. 
Only the largest values are shown, taken from the bins
with the largest effect when the systematic variation was applied.
}
\label{tab_system}
\end{table*}

\section{Results}

\begin{figure*}
   \begin{center}
   \includegraphics[width=15cm]{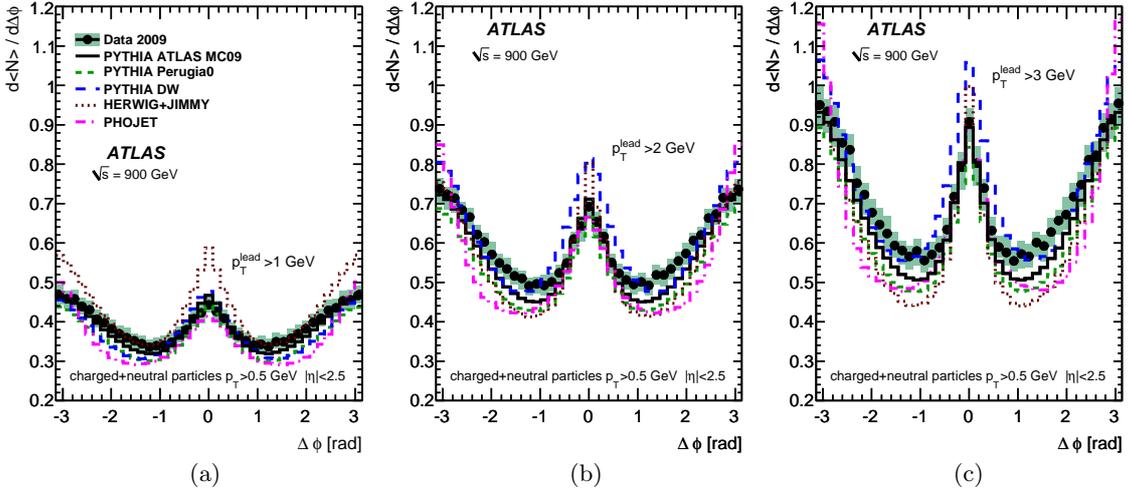}
   \begin{picture} (0.,0.)
   \setlength{\unitlength}{1.0cm}

   \put (-12.6,-0.05){(a)}
   \put (-7.6,-0.05){(b)} 
   \put (-2.55,-0.05){(c)}

 \end{picture}
\vspace*{0.5cm}
   \caption{The average number of particles per unit of pseudorapidity as a function of the
            azimuthal separation between the leading particle and other particles in
            inelastic $pp$ collisions at
            $\sqrt{s}=900$~\GeV. The shaded band shows the statistical and systematic
            uncertainties added in quadrature. The
            densities are shown for (a) $\ptl>1$~\GeV, (b) $\ptl>2$~\GeV and (c) $\ptl>3$~\GeV.}
    \label{fig:final_cluster_phiphi}
  \end{center}
\end{figure*}

Figure~\ref{fig:final_cluster_phiphi}
shows the density distribution  $d\langle N \rangle /d \Delta \phi$ of stable-particles 
as a function
of the distance in azimuthal angle between the leading particle
and other particles in an event for $\sqrt{s}=900$~\GeV.  
This density, defined in Sect.~\ref{corr}, is calculated
for events selected by requiring a particle with
$\ptl$ above the values indicated on the figure.
The detector correction for this density is discussed in Sect.~\ref{corr}. 
The total uncertainty, computed from the addition of statistical and systematic 
uncertainties in quadrature, is shown as a shaded band on all measurements.

\begin{figure*}
   \begin{center}
   \includegraphics[width=15cm]{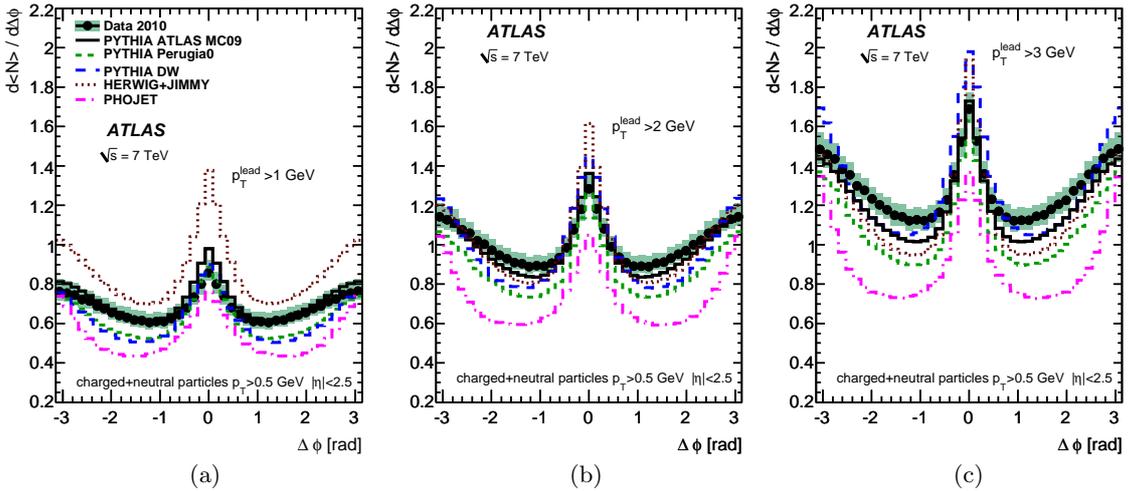}
 \begin{picture} (0.,0.)
   \setlength{\unitlength}{1.0cm}

   \put (-12.6,-0.05){(a)}
   \put (-7.6,-0.05){(b)}
   \put (-2.55,-0.05){(c)}

 \end{picture}
\vspace*{0.5cm}
   \caption{
            The average number of particles per unit of pseudorapidity as a function of the
            azimuthal separation between the leading particle and other particles in
            inelastic $pp$ collisions at
            $\sqrt{s}=7$~\TeV. The shaded band shows the statistical and systematic
            uncertainties added in quadrature. The
            densities are shown for (a) $\ptl>1$~\GeV, (b) $\ptl>2$~\GeV and (c) $\ptl>3$~\GeV.
            }
    \label{fig:final_cluster_phiphi_7tev}
  \end{center}
\end{figure*}

The angular distribution shown in Fig.~\ref{fig:final_cluster_phiphi} has a 
peak at $\Delta\phi\simeq 0$ which reflects the particle activity from the hard interaction. 
The  peak narrows as $\ptl$ increases. 
The shape of this distribution is similar to that observed 
in the recent track-based publications~\cite{CDF:mbpaper,dd:2010rm, CMS:UE,ATLAS:UE},
and also similar to the transverse-momentum flow around jets  
observed at a lower $p\bar{p}$ collision energy~\cite{Albajar1988405}. 
The particle densities measured using topological clusters are
higher than the charged-particle densities measured using tracks \cite{ATLAS:UE}, which is 
expected from the neutral-particle contribution.

Figure~\ref{fig:final_cluster_phiphi_7tev}
shows the $\Delta\phi$ density distributions 
for $\sqrt{s}=7$~\TeV. 
The distributions show narrower peaks, for
a given $\ptl$ threshold, than for the $\sqrt{s}=900$~\GeV data.

\begin{figure*}[ht]
\begin{minipage}[b]{0.5\linewidth}
\centering
\includegraphics[scale=0.38]{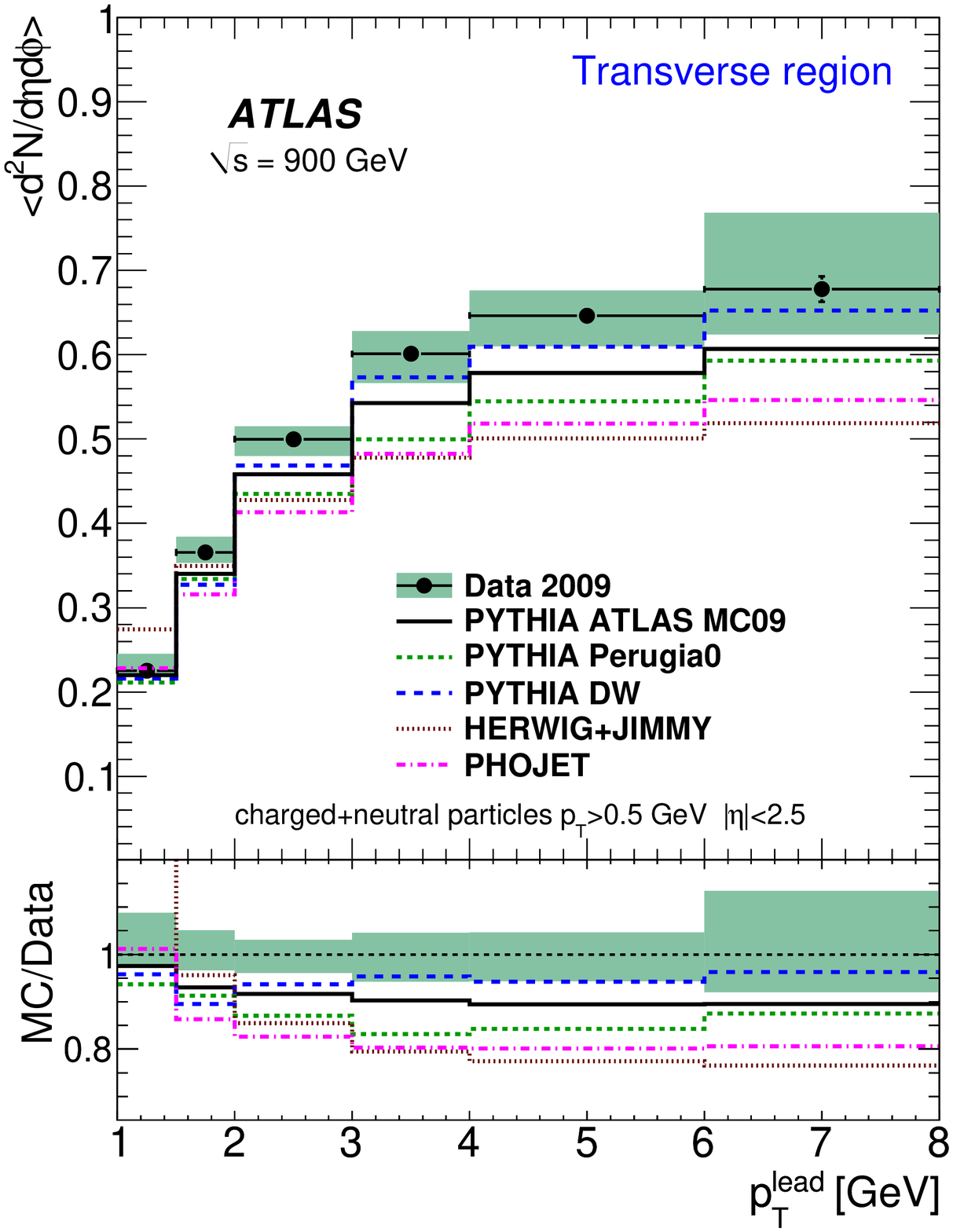}
\end{minipage}
\hspace{0.3cm}
\begin{minipage}[b]{0.5\linewidth}
\centering
\includegraphics[scale=0.38]{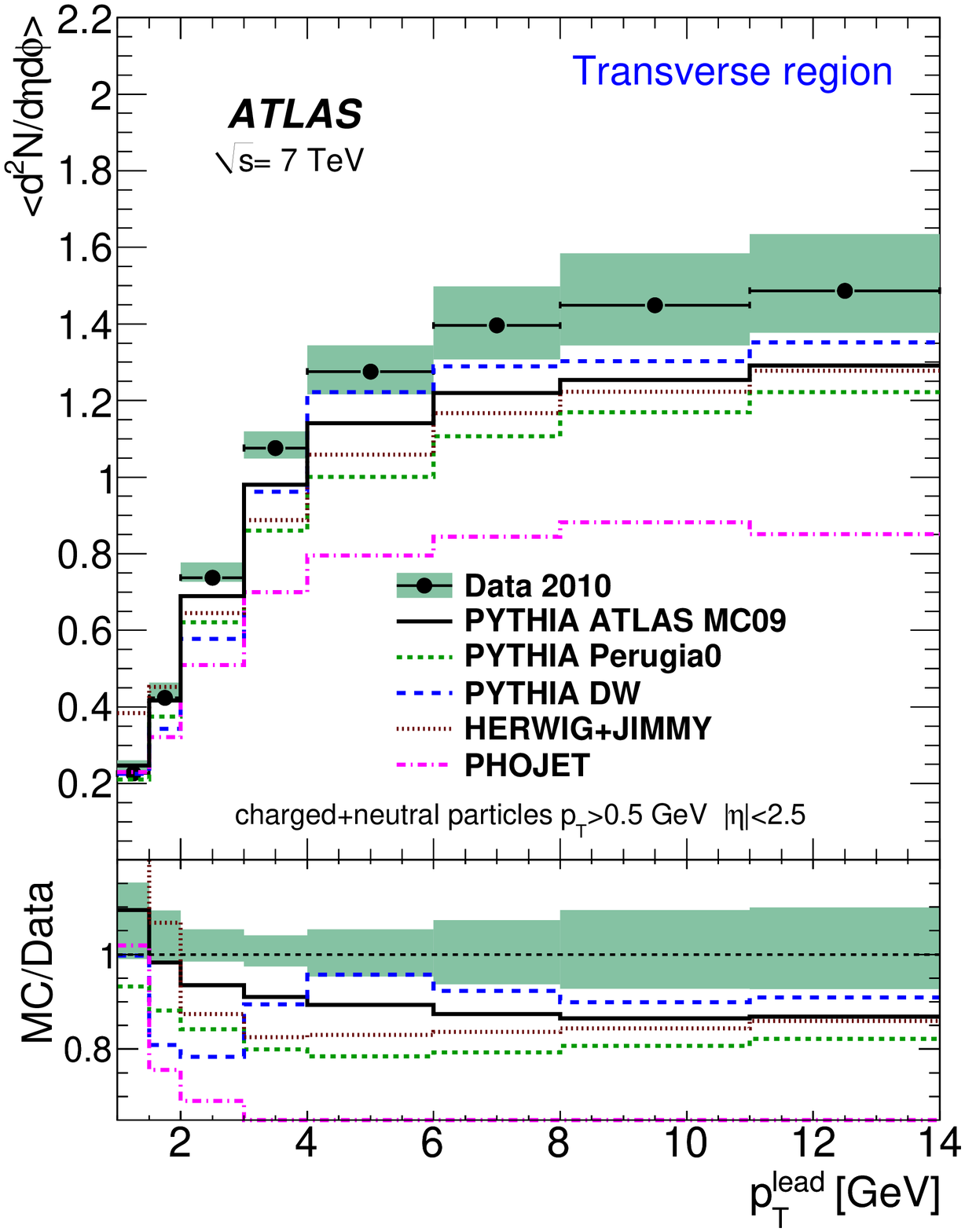}
\end{minipage}

\begin{picture} (0.,0.)
   \setlength{\unitlength}{1.0cm}

   \put (4.5,0.3){(a)}
   \put (14.4,0.3){(b)}

 \end{picture}
\vspace*{0.2cm}
\caption{
The average number of stable particles per event per unit interval in $\eta$--$\phi$,
as a function of $\ptl$, for the transverse region
indicated in Fig.~\protect\ref{fig:uediagram}.
The results are shown for (a) $\sqrt{s}=900$~\GeV and (b) for $\sqrt{s}=7$~\TeV $pp$ collisions.
The shaded band shows the statistical and systematic
uncertainties added in quadrature.
}
\label{fig:final_cluster_ptlead_trans}
\end{figure*}

The data are compared to the PYTHIA Monte Carlo generator with the MC09, Perugia0 and DW
tunes, PHOJET and HERWIG+JIMMY.  
The Monte Carlo generators reproduce the general
features of the data, but fail to describe the detailed behaviour, as can
be seen in the figures. 
The MC09 and Perugia0 PYTHIA tunes are closest to the data.
The PHOJET generator significantly underestimates the particle densities,
while the PYTHIA DW and HERWIG overestimate the data at $\Delta\phi\simeq$0.
The data are seen to have a large discriminating
power and are thus useful to constrain the parameters of Monte Carlo
generators.

Figure~\ref{fig:final_cluster_ptlead_trans}
shows the mean number of particles per event per unit interval in $\eta$ 
and $\phi$ as defined in Sect.~\ref{corr}. 
The density was calculated in the transverse region illustrated in Fig.~\protect\ref{fig:uediagram},
as a function of $\ptl$. None of the Monte Carlo predictions describe the data 
well. The DW tune is the most similar to the observed data.  As is seen in the $\Delta\phi$
distribution, the PHOJET simulation lacks a hard component for $\sqrt{s}=7$~\TeV. 
The particle density increases almost by a factor of two,  going from $\sqrt{s}=900$~\GeV to 
$\sqrt{s}=7$~\TeV at a similar $\ptl$, which is comparable to what is seen in all Monte Carlo
generators.

\begin{figure*}[ht]
\begin{minipage}[b]{0.5\linewidth}
\centering
\includegraphics[scale=0.38]{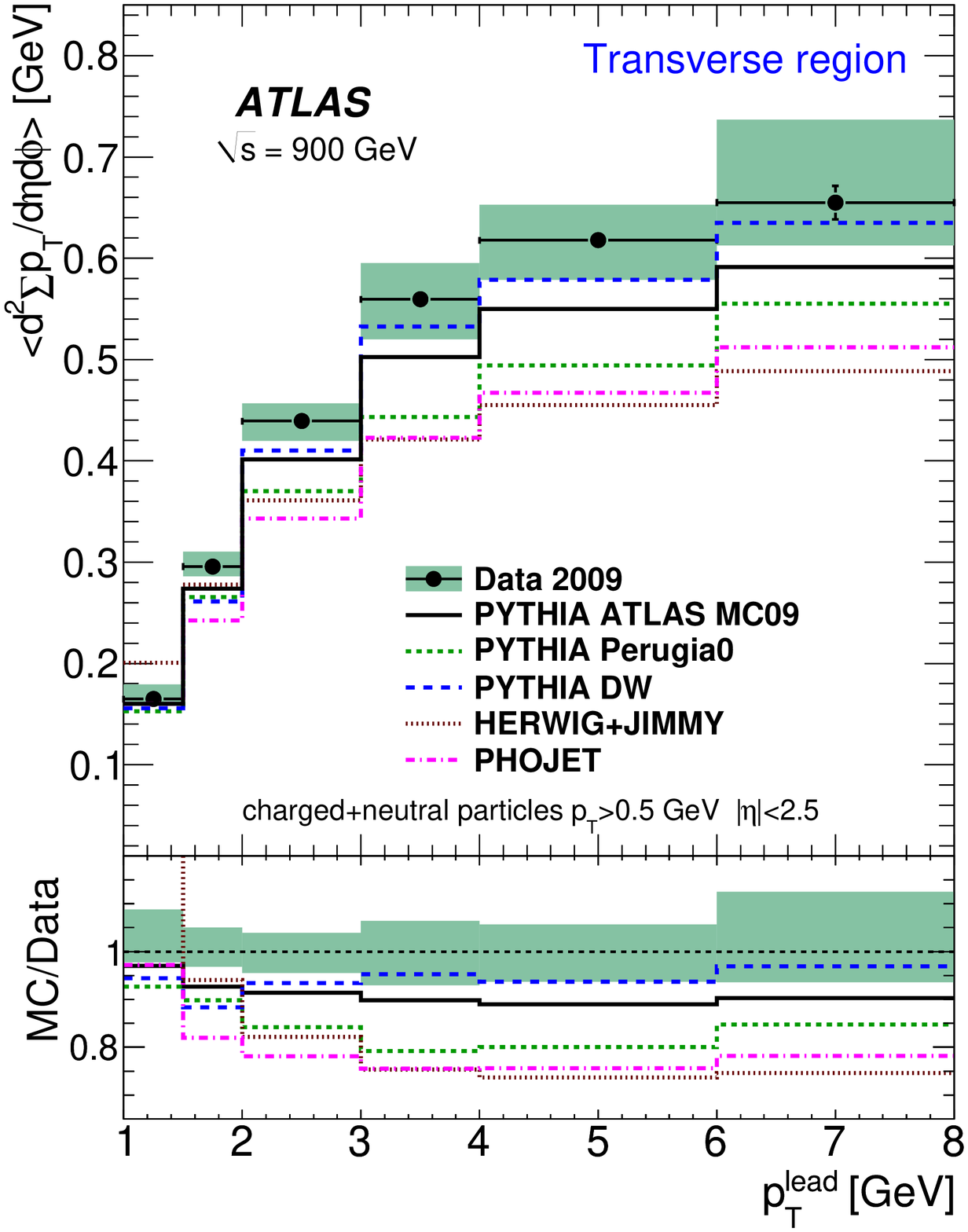}
\end{minipage}
\hspace{0.3cm}
\begin{minipage}[b]{0.5\linewidth}
\centering
\includegraphics[scale=0.38]{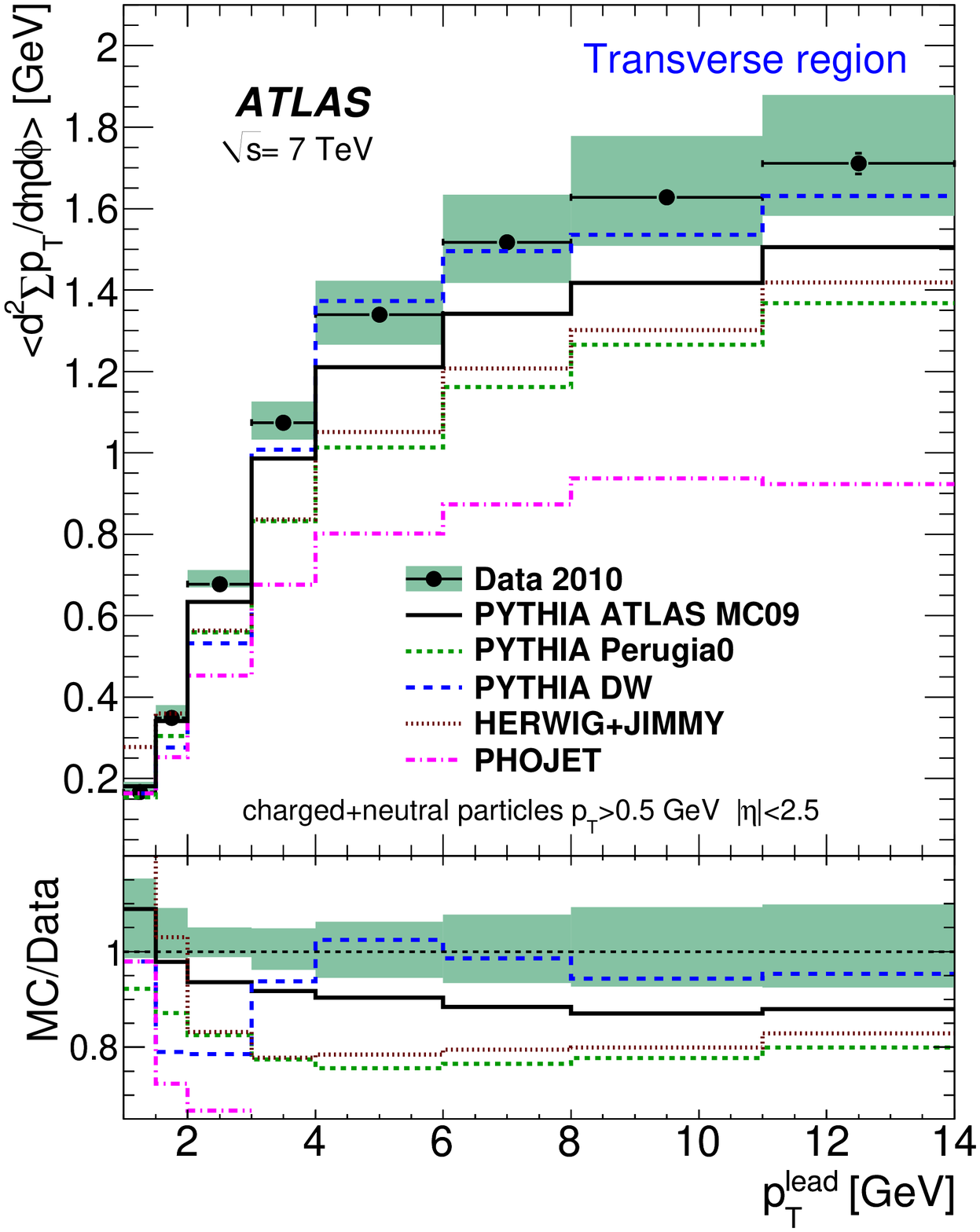}
\end{minipage}
\begin{picture} (0.,0.)
   \setlength{\unitlength}{1.0cm}

   \put (4.5,0.3){(a)}
   \put (14.4,0.3){(b)}

 \end{picture}
\vspace*{0.2cm}
\caption{
The average scalar $\pT$ sum for stable particles per unit area in $\eta$--$\phi$ in the transverse region as a
function of $\ptl$ for (a) $\sqrt{s}=900$~\GeV  and (b)  $\sqrt{s}=7$~\TeV data. 
The shaded band shows the statistical and systematic
uncertainties added in quadrature.
}
\label{fig:final_cluster_ptavlead_trans}
\end{figure*}

Figure~\ref{fig:final_cluster_ptavlead_trans}
shows the mean scalar $\pT$ sum for stable particles in the transverse region
as a function of $\ptl$. As for the particle densities, 
the mean transverse-momentum  sum
is measured per unit interval in $\eta$ and $\phi$ (see Sect.~\ref{corr}).   
Again, the Monte Carlo predictions do not fully describe the data.
The largest discrepancy with the data is found for the PHOJET generator.

\newpage
\section{Conclusions}

Particle densities sensitive to the underlying event
in $pp$ collisions at centre-of-mass energies of 900 \GeV and 7 \TeV \\ are presented. 
This is the first such analysis completely based  
on calorimetric measurement of three-dimensional energy depositions,
which is made possible by the fine granularity of the ATLAS calorimeter with
transverse and longitudinal samplings.

The particle densities 
were studied and compared with several Monte Carlo generators
tuned to pre-LHC data. 
None of the Monte Carlo generators describe the measurements well.
In particular, the Monte Carlo predictions have discrepancies with the data  
for the particle density as a  function of the azimuthal
angle between the leading particle and any other particle in an event.
The Monte Carlo generators typically predict a lower particle density in the
transverse region ($|\Delta\phi|\simeq \pi/2$), while in the toward region
($\Delta\phi\simeq0$), the PYTHIA DW and HERWIG+JIMMY generators both overestimate 
the densities. 
PHOJET significantly fails for the $\sqrt{s}=7$~\TeV data.
For the particle densities as a function of $\ptl$,
all the Monte Carlo generators also fail to describe the data, 
predicting lower than observed particle activity in the
transverse region. A similar conclusion holds for the total transverse momentum
of particles in the transverse region.

The particle densities measured using topological clusters are 
higher than the charged-particle densities measured using tracks~\cite{ATLAS:UE}.
This is expected from the neutral-particle contribution. 
The discrepancies between the data  and 
Monte Carlo generators agree with those observed for charged
particles \cite{ATLAS:UE}.
These measurements have systematic uncertainties independent to the
track-based studies and provide additional information
which may be used to
improve the Monte Carlo description
of the complete
final state produced in $pp$ collisions.

\section*{Acknowledgements}

We wish to thank CERN for the efficient commissioning and operation of the LHC during this initial high-energy data-taking period as well as the support staff from our institutions without whom ATLAS could not be operated efficiently.

We acknowledge the support of ANPCyT, Argentina; YerPhI, Armenia; ARC, Australia; BMWF, Austria; ANAS, Azerbaijan; SSTC, Belarus; CNPq and FAPESP, Brazil; NSERC, NRC and CFI, Canada; CERN; CONICYT, Chile; CAS, MOST and NSFC, China; COLCIENCIAS, Colombia; MSMT CR, MPO CR and VSC CR, Czech Republic; DNRF, DNSRC and Lundbeck Foundation, Denmark; ARTEMIS, European Union; IN2P3-CNRS, \\ CEA-DSM/IRFU, France; GNAS, Georgia; BMBF, DFG, HGF, MPG and AvH Foundation, Germany; GSRT, Greece; ISF, MINERVA, GIF, DIP and Benoziyo Center, Israel; INFN, Italy; MEXT and JSPS, Japan; CNRST, Morocco; FOM and NWO, Netherlands; RCN, Norway;  MNiSW, Poland; GRICES and FCT, Portugal;  MERYS (MECTS), Romania;  MES of Russia and ROSATOM, Russian Federation; JINR; MSTD, Serbia; MSSR, Slovakia; ARRS and MVZT, Slovenia; DST/NRF, South Africa; MICINN, Spain; SRC and Wallenberg Foundation, Sweden; SER,  SNSF and Cantons of Bern and Geneva, Switzerland;  NSC, Taiwan; TAEK, Turkey; STFC, the Royal Society and Leverhulme Trust, United Kingdom; DOE and NSF, United States of America.

The crucial computing support from all WLCG partners is acknowledged gratefully, in particular from CERN and the ATLAS Tier-1 facilities at TRIUMF (Canada), NDGF (Denmark, Norway, Sweden), CC-IN2P3 (France), KIT/GridKA (Germany), INFN-CNAF (Italy), NL-T1 \\ (Netherlands), PIC (Spain), ASGC (Taiwan), RAL (UK) and BNL (USA) and in the Tier-2 facilities worldwide.

\bibliographystyle{l4z_epj}
\bibliography{ueclusters_paper}

\providecommand{\etal}{et al.\xspace}
\providecommand{\coll}{Collab.\xspace}
\catcode`\@=11
\def\@bibitem#1{%
\ifmc@bstsupport
  \mc@iftail{#1}%
    {;\newline\ignorespaces}%
    {\ifmc@first\else.\fi\orig@bibitem{#1}}
  \mc@firstfalse
\else
  \mc@iftail{#1}%
    {\ignorespaces}%
    {\orig@bibitem{#1}}%
\fi}%
\catcode`\@=12
\begin{mcbibliography}{10}

\bibitem{r.field}
R.~Field,
\newblock Acta~Phys.~Pol.{} {\bf B~36},~167~(2005)\relax
\relax
\bibitem{CDF:mbpaper}
{The CDF}~Collaboration, T.~Aaltonen, et~al.,
\newblock Phys.~Rev.~D{} {\bf 70},~072002~(2004)\relax
\relax
\bibitem{dd:2010rm}
{The CDF}~Collaboration, T.~Aaltonen, et~al.,
\newblock Phys.~Rev.{} {\bf D~82},~034001~(2010)\relax
\relax
\bibitem{CMS:UE}
{The CMS}~Collaboration, V.~Khachatryan, et~al.,
\newblock Eur.~Phys.~J{} {\bf C~70},~555~(2010)\relax
\relax
\bibitem{ATLAS:UE}
{The ATLAS Collaboration},
\newblock Phys.~Rev.~D (in press){}~(2010)\relax
\relax
\bibitem{ATLAS:detector}
{The ATLAS Collaboration},
\newblock JINST{} {\bf 3},~S08003~(2008)\relax
\relax
\bibitem{Lampl:1099735}
{The ATLAS Collaboration},
\newblock ATLAS-LARG-PUB-2008-002{}~(2008)\relax
\relax
\bibitem{Barillari:1112035}
{The ATLAS Collaboration},
\newblock ATLAS-LARG-PUB-2009-001{}~(2008)\relax
\relax
\bibitem{CALOR:H1calib}
B.~Andrieu, et~al.,
\newblock Nucl.~Instrum.~Meth.{} {\bf A~386},~348~(1997)\relax
\relax
\bibitem{Abat:2009zz}
E.~Abat, E.~Arik, S.~Cetin, J.~Abdallah, M.~Bosman, et~al.,
\newblock Nucl.~Instrum.~Meth.{} {\bf A~607},~372~(2009)\relax
\relax
\bibitem{Adragna:2009zz}
P.~Adragna, C.~Alexa, K.~Anderson, A.~Antonaki, A.~Arabidze, et~al.,
\newblock Nucl.~Instrum.~Meth.{} {\bf A~606},~362~(2009)\relax
\relax
\bibitem{Aharrouche:2010zz}
M.~Aharrouche, C.~Adam-Bourdarios, M.~Aleksa, D.~Banfi, D.~Benchekroun, et~al.,
\newblock Nucl.~Instrum.~Meth.{} {\bf A~614},~400~(2010)\relax
\relax
\bibitem{2009zp}
{The ATLAS Collaboration},
\newblock Eur.~Phys.~J.{} {\bf C~70},~723~(2010)\relax
\relax
\bibitem{2010dd}
{The ATLAS Collaboration},
\newblock Eur.~Phys.~J.{} {\bf C~70},~1193~(2010)\relax
\relax
\bibitem{Aad:2010rd}
{The ATLAS Collaboration},
\newblock Phys.~Lett.{} {\bf B~688},~21~(2010)\relax
\relax
\bibitem{ATLAS-CONF-2010-069}
{The ATLAS Collaboration},
\newblock ATLAS-CONF-2010-069{}~(2010)\relax
\relax
\bibitem{Sjostrand:2006za}
T.~Sjostrand, S.~Mrenna, P.~Z. Skands,
\newblock JHEP{} {\bf 05},~026~(2006)\relax
\relax
\bibitem{ATL-PHYS-PUB-2010-002}
{The ATLAS Collaboration},
\newblock ATLAS-PHYS-PUB-2010-002{}~(2010)\relax
\relax
\bibitem{Sherstnev:2007nd}
A.~Sherstnev, R.~S. Thorne,
\newblock Eur.~Phys.~J.{} {\bf C~55},~553~(2008)\relax
\relax
\bibitem{Andersson:1983ia}
B.~Andersson, G.~Gustafson, G.~Ingelman, T.~Sjostrand,
\newblock Phys.~Rept.{} {\bf 97},~31~(1983)\relax
\relax
\bibitem{Abe:1988yu}
{The CDF}~Collaboration, F.~Abe, et~al.,
\newblock Phys.~Rev.~Lett.{} {\bf 61},~1819~(1988)\relax
\relax
\bibitem{Skands:2009zm}
P.~Z. Skands~(2009).
\newblock {Presented at Multi-Parton Interaction Workshop, Perugia, Italy,
  28-31 Oct 2008. Published in Perugia 2008, Multiple Partonic Interactions at
  the LHC (MPI08) p.284}\relax
\relax
\bibitem{DW}
R.~Field,
\newblock {\em {MinBias and the Underlying Event at the Tevatron and the LHC, A
  talk presented at the Fermilab mc tuning workshop}},
\newblock Technical Report, FERMILAB, 2002\relax
\relax
\bibitem{Engel:1999xb}
R.~Engel, J.~Ranft,
\newblock Nucl.~Phys.~Proc.~Suppl.{} {\bf A~75},~272~(1999)\relax
\relax
\bibitem{Corcella:2000bw}
G.~Corcella, et~al.,
\newblock JHEP{} {\bf 01},~010~(2001)\relax
\relax
\bibitem{Corcella:2002jc}
G.~Corcella, et~al.,
\newblock {\em {HERWIG 6.5 release note}}.
\newblock ArXiv:0210213, 2002\relax
\relax
\bibitem{Butterworth:1996zw}
J.~M. Butterworth, J.~R. Forshaw, M.~H. Seymour,
\newblock Z.~Phys.{} {\bf C~72},~637~(1996)\relax
\relax
\bibitem{Marshall:1136205}
{The ATLAS Collaboration},
\newblock Nucl.~Phys.~Proc.~Suppl.{} {\bf 197},~254~(2009)\relax
\relax
\bibitem{Agostinelli2003250}
S.~Agostinelli, et~al.,
\newblock Nucl.~Instr.~Meth.{} {\bf A~506},~250~(2003)\relax
\relax
\bibitem{Barberio:1271712}
{The ATLAS Collaboration},
\newblock ATLAS-CONF-2010-052{}~(2010)\relax
\relax
\bibitem{ATLAS-CONF-2010-017}
{The ATLAS Collaboration},
\newblock ATLAS-CONF-2010-017{}~(2010)\relax
\relax
\bibitem{ATLAS-CONF-2010-053}
{The ATLAS Collaboration},
\newblock {ATLAS-CONF-2010-053}{}~(2010)\relax
\relax
\bibitem{Cacciari:2008gp}
M.~Cacciari, G.~P. Salam, G.~Soyez,
\newblock JHEP{} {\bf 04},~063~(2008)\relax
\relax
\bibitem{Albajar1988405}
UA1~Collaboration, C.~Albajar, et~al.,
\newblock Nucl.~Phys.{} {\bf B~309},~405~(1988)\relax
\relax
\end{mcbibliography}

\newpage

\begin{flushleft}
{\Large The ATLAS Collaboration}

\bigskip

G.~Aad$^{\rm 48}$,
B.~Abbott$^{\rm 111}$,
J.~Abdallah$^{\rm 11}$,
A.A.~Abdelalim$^{\rm 49}$,
A.~Abdesselam$^{\rm 118}$,
O.~Abdinov$^{\rm 10}$,
B.~Abi$^{\rm 112}$,
M.~Abolins$^{\rm 88}$,
H.~Abramowicz$^{\rm 153}$,
H.~Abreu$^{\rm 115}$
% ...
\bigskip

% ...
$^{10}$ Institute of Physics, Azerbaijan Academy of Sciences, Baku, Azerbaijan\\
$^{11}$ Institut de F\'isica d'Altes Energies and Universitat Aut\`onoma  de Barcelona and ICREA, Barcelona, Spain\\
% ...
$^{48}$ Fakult\"{a}t f\"{u}r Mathematik und Physik, Albert-Ludwigs-Universit\"{a}t, Freiburg i.Br., Germany\\
$^{49}$ Section de Physique, Universit\'e de Gen\`eve, Geneva, Switzerland\\
% ...
$^{88}$ Department of Physics and Astronomy, Michigan State University, East Lansing MI, United States of America\\
% ...
$^{111}$ Homer L. Dodge Department of Physics and Astronomy, University of Oklahoma, Norman OK, United States of America\\
$^{112}$ Department of Physics, Oklahoma State University, Stillwater OK, United States of America\\
% ...
$^{115}$ LAL, Univ. Paris-Sud and CNRS/IN2P3, Orsay, France\\
% ...
$^{118}$ Department of Physics, Oxford University, Oxford, United Kingdom\\
% ...
$^{153}$ Raymond and Beverly Sackler School of Physics and Astronomy, Tel Aviv University, Tel Aviv, Israel\\
% ...
\end{flushleft}

\end{document}